\documentclass[aps,prl,prabib,epsfig,amsmath,amssymb,twocolumn,superscriptaddress,showpacs]{revtex4}
\usepackage{amsmath}
\usepackage{amssymb}
\usepackage{epsfig}
\usepackage{graphics}

\newcommand \beq{\begin{eqnarray}}
\newcommand \eeq{\end{eqnarray}}
\def\simge{\mathrel{%
       \rlap{\raise 0.511ex \hbox{$>$}}{\lower 0.511ex \hbox{$\sim$}}}}
\def\simle{\mathrel{
       \rlap{\raise 0.511ex \hbox{$<$}}{\lower 0.511ex \hbox{$\sim$}}}}
\newcommand{\qq}{\mathbf{q}}
\newcommand{\pp}{\mathbf{p}}

\begin{document}

\title{Comment on ``Normal phase of an imbalanced Fermi gas"}

\author{Zhenhua Yu}
\affiliation{Niels Bohr International Academy, Niels Bohr Institute,  DK-2100 Copenhagen \O, Denmark}
\author{Sascha Z\"ollner}
\affiliation{Niels Bohr International Academy, Niels Bohr Institute,  DK-2100 Copenhagen \O, Denmark}
\author{C.\ J.\  Pethick}
\affiliation{Niels Bohr International Academy, Niels Bohr Institute,  DK-2100 Copenhagen \O, Denmark}
\affiliation{ NORDITA, Roslagstullsbacken 21, SE-10691 Stockholm, Sweden}

\maketitle\vspace{-1em}

Recently Mora and Chevy \cite{mora}, in studying the energy of an atomic Fermi gas consisting of a majority species, 1, and a minority species, 2, showed that, due to interatomic interactions, the energy density $E$ of the gas  has a contribution of the form
$E^{(2)}= \frac12  f  n_2^2$, where $n_i$ is the density of species $i$.   They  attribute this term to a ``modification of the single polaron
properties due to Pauli blocking'. In this Comment, we demonstrate that $E^{(2)}$ may equivalently be understood in terms of the familiar interaction between minority atoms induced by the majority component.  Our treatment is inspired by the theory of dilute solutions of $^3$He in superfluid $^4$He \cite{BBP}.  

We consider the interaction between  two minority atoms in momentum states $\pp$ and $\pp'$, since $f$ is the Landau quasiparticle interaction between minority atoms. Since the momenta are small compared with other relevant momentum scales,  we neglect the momentum dependence of the interaction.
For the moment, we regard the two atoms as being distinguishable, and the effects of quantum statistics will be taken into account later.  The effective interaction, $T_{\pp \pp'}$, between the atoms is the change in the energy of a quasiparticle in the momentum state $\pp$ when the distribution function for quasiparticles in the state $\pp'$ is varied, and it may be written in the form \cite{BBP}
\beq
T_{\pp \pp'}= \left.T_{\pp \pp'}\right|_{\rm dir} + \left.T_{\pp \pp'}\right|_{\rm ind}.
\label{T}
\eeq
The first term, which does not involve changes in the distribution function for the majority atoms, is the direct interaction and the second term, which takes into account interactions mediated by the majority atoms, is the induced interaction.  The latter has the standard form
\beq
\left.T_{\pp \pp'}\right|_{\rm ind}=  g_{12}^ 2 \chi(\qq,\omega).
\eeq
Here $g_{12}$ is the effective interaction between a majority atom and a minority one, given by the energy change of a majority atom when a minority one is added,  $g_{12}=\partial^2 E/\partial n_1\partial n_2=\partial \mu_2/\partial n_1$, where $\mu_i=\partial E/\partial n_i$ is the chemical potential of species $i$.  The quantity  $\chi(\qq,\omega)=-\partial n_1/\partial \mu_1\{1-(s/2) \ln[(s+1)/(s-1)]\}$ is the density--density response function  for noninteracting majority atoms. Here $s=\omega/v_F q$,  with $\omega$ the frequency and $q$ the wavenumber at which the quasiparticle distribution is modulated, and $v_F$ is the Fermi velocity of the majority atoms. Interactions have a negligible effect on the response function for low densities $n_1$. Since s-wave interactions between majority atoms vanish as a consequence of the Pauli principle, interactions between majority atoms are unimportant, and those between majority and minority atoms can be neglected since $n_2$ is small.

Finally we evaluate  the Landau quasiparticle interaction from the effective interaction.  If the minority atoms were distinguishable, there would be no contribution from the induced interaction, since Landau parameters are obtained from effective interactions by taking first the limit $q\to 0$ and then the limit 
$\omega \to 0$ \cite{AGD}, and $\chi$ vanishes in this limit.  However, for identical fermions, the total contribution to the Landau interaction consists of the sum of expression (\ref{T}) and the corresponding exchange term. The induced interaction contribution to the exchange term does not vanish in this limit, because the momentum transfer is equal to $\pp-\pp'$, and therefore remains nonzero under the limiting process.   Because of the Pauli principle, the contribution from the direct term is zero, and therefore the Landau quasiparticle interaction is given by
\vspace{-1em}
\beq
f=\left(\frac{\partial\mu_2}{\partial n_1}\right)^2\frac{\partial n_1}{\partial \mu_1}=\nu^2\frac{\partial \mu_1}{\partial n_1},
\label{f}
\eeq
where $\nu=-(\partial \mu_2/\partial n_1)/(\partial \mu_1/\partial n_1)$.  Physically, $\nu=dn_2/dn_1|_{\mu_1}$ is the number of majority atoms in the dressing cloud of a minority atom.  Even though the induced interaction is intrinsically attractive at low frequency, $f$ is positive because it comes from exchange. The interaction contribution to the energy per unit volume is therefore $\nu^2({\partial \mu_1}/{\partial n_1})n_2^2/2$, in agreement with   Eqs.\ (1) and (3) of Ref.\ \cite{mora}.  

To summarize, the density-mediated (induced) interaction accounts entirely for the result of Ref.\ \cite{mora} for the effective interaction between minority atoms.  

We are grateful to Gordon Baym for helpful comments.  SZ was supported by the German Academy of Sciences Leopoldina.

\end{document}